\documentclass[11pt]{article}
\usepackage[dvips]{color}
\usepackage{epsfig}
\usepackage{amsmath}
\usepackage{graphicx}
\date{}
\begin{document}
\title{{\bf Polymer deformation and particle tunneling from Schwarzschild black hole}}
\author{Majid Amirfakhrian$^1$\thanks{%
e-mail: m5764amirfakhrian@gmail.com}\,\, and Babak
Vakili$^{2,3}$\thanks{email: b.vakili@iauctb.ac.ir}
\\\\
$^1${\small {\it Department of Physics, Science and Research Branch,
Islamic Azad University, Tehran, Iran}}\\
$^2${\small {\it Department of Physics, Central Tehran Branch,
Islamic Azad University, Tehran, Iran}}\\$^3$ {\small {\it Research
Institute for Astronomy and Astrophysics of Maragha (RIAAM)-
Maragha, IRAN, P. O. Box: 55134-441}}} \maketitle
\begin{abstract}
In this paper, we investigate a tunneling mechanism of massless
particles from the Schwarzschild black hole in the framework of
polymer quantum mechanics. According to the corresponding invariant
Liouville volume, we determine the tunneling rate from Schwarzschild
black hole by the polymeric quantization procedure. In this regard,
we show that the temperature and tunneling radiation of the black
hole receive new corrections in such a way that the exact radiant
spectrum is no longer precisely
thermal.\vspace{5mm}\noindent
\end{abstract}

\section{Introduction}

From the time that for the first time the minimal length idea in
presence of quantum gravity was propounded, lots of endeavors have
been made in this field and remarkable works has been introduced
\cite{Veneziano86}. On this basis, the uncertainty principle of
Heisenberg was generalized by fundamental measurable length which
today it is known as generalized uncertainty principles (GUP)
\cite{Meissner04}. Some of the phenomenological concepts of GUP
which have been studied for example in \cite{Camelia01,ADK,Beh},
show a meaningful compatibility with the other approaches of quantum
gravity \cite{rov}. Especially in \cite {Beh} a method is proposed
to test the effects of quantum gravity on the thermodynamical
properties of a sonic black hole or dumb hole. By this method the
usual logarithmic correction to the entropy is generated with
correct coefficient. As an inspiration from the primitive attempts
of Snyder in 1947 who formulated the Lorentz invariant discrete
space-time, a symplectic structure of phase space in none-canonical
framework is formulated in the non-relativistic limit \cite{snyder}.
The results of the deformed phase space is very similar to the
string theory which leads us to a new definition of GUP
\cite{snyder2}. From a geometrical point of view, the authors in
\cite{Cap} have shown that GUP may be extracted from quantum
geometry in which the quantum effects are encoded in the space-time
geometry. They did this by considering an upper limit on the
acceleration of massive particles and concluded that it affects the
canonical commutation relations, deform them in the form of GUP.

Between the useful notions that also use a minimal length scale in
their formalism, we can mention the polymer quantization
\cite{Corichi1}, which uses the methods very similar to the
effective theories of loop quantum gravity. In polymer approach of
quantum mechanics a polymer length scale, $\lambda$, which shows the
scale of the segments of the granular space enters into the
Hamiltonian of the system to deform its functional form into a
so-called polymeric Hamiltonian. This means that in a polymeric
quantized system in addition of a quantum parameter $\hbar$, which
is responsible to canonical quantization of the system, there is
also another quantum parameter $\lambda$, that labels the granular
properties of the underlying space. This approach then opened a new
window for the theories which are dealing with the quantum
gravitational effects in physical systems such as quantum cosmology
and black hole physics, see for instance \cite{qc pol} and \cite{pol
bl} and the references therein.

Recently, Parikh and Wilzcek have shown a smooth model for
performing the semi-classical tunneling process from the event
horizon of black holes (BH) by correcting the radiation of
self-gravitation and incorporating the modified background geometry
as a dynamical system \cite{parikh}. Following this idea, though a
number of recent papers have tried to extend this semi-classical
tunneling method for various types of BHs such as rotating and
charged BHs \cite{Vag01}, less research is focused in high energy
scales.

In this letter, incited by the above expressed, we want
to investigate how the tunneling mechanism from the Schwarzschild
black hole (S-BH) gets modifications due to the polymerization.
Since the temperature of a BH can be written in terms of the
tunneling rate, the corrections to the BH's tunneling rate yield
naturally modifications to its thermodynamics. To do this, we begin
with a general form of a S-BH space-time.  We then follow the
Parikh-Wilczek procedure to evaluate the polymer corrected tunneling
probability. So, after a brief review of the polymer representation
and classical polymerization in section 2, we shall deal with the
tunneling mechanism in polymer phase space in section 3. We
summarize the results in section 4.

\section{Polymerization}\label{sec2}
In polymeric formulation of quantum mechanics, the position space
(labeled by the coordinate $q$) is presumed to be discrete. Indeed,
this space consists of small segments with discreteness parameter
$\lambda$. This makes it impossible to define the generator of the
displacement in a natural way and thus the momentum operator
associated to $q$ does not exist \cite{QPR}. Nevertheless, it is
possible to find an effective momentum in the semi-classical picture
by means of the Weyl exponential operator (shift operator). To do
this, let us consider a function $f(q)$ which its derivative with
respect to the discrete coordinate $q$ may be estimated by utilizing
the Weyl operator as \cite{CPR}

\begin{eqnarray}\label{FWD}
\partial_{q}f(q)\approx\frac{1}{2\lambda}[f(q+\lambda)-f(q-
\lambda)]\hspace{2cm}\nonumber\\=\frac{1}{2\lambda}\Big(
\widehat{e^{ip\lambda}}-\widehat{e^{-ip\lambda}}\Big)\,f(q)=
\frac{i}{\lambda}\widehat{\sin(\lambda p)}\,f(q),
\end{eqnarray}
and in a similar way, the second derivative will be
\begin{eqnarray}\label{SWD}
\partial_{q}^2f(q)\approx\frac{1}{\lambda^2}[f(q+\lambda)-2
f(q)+f(q-\lambda)]\hspace{1cm}\nonumber\\=\frac{2}{\lambda^2}
(\widehat{\cos(\lambda p)}-1)\,f(q).\hspace{2cm}
\end{eqnarray}
Having the above approximations at hand, we define the
polymerization process for the finite values of the parameter
$\lambda$ as

\begin{eqnarray}\label{Polymerization}
\hat{p}\rightarrow\,\frac{1}{\lambda}\widehat{\sin(\lambda p)},
\hspace{1cm}\hat{p}^2\rightarrow\,\frac{2}{\lambda^2}(1-
\widehat{\cos(\lambda p)}).
\end{eqnarray}
One may extend these substitutions to the classical dynamical
variables so that a classical theory is achieved through above
mechanism but now without any attribution to the Weyl operator. This
is what usually is called classical polymerization in the literature
of quantum gravity \cite{Corichi1}, \cite{CPR}

\begin{eqnarray}\label{PT}
q\rightarrow q,\hspace{1.5cm}p\rightarrow\frac{ \sin(\lambda
p)}{\lambda},\hspace{5mm}p^2\rightarrow
\frac{2}{\lambda^2}\left[1-\cos(\lambda p)\right],
\end{eqnarray}
where the pair $(p,q)$ are variables of classical phase space. By
means of this approach, a polymerized classical system can be
described by the Hamiltonian

\begin{equation}\label{PHamiltonian}
H_{\lambda}=\frac{1}{m\lambda^2}\big(1-\cos(\lambda p)\big) +U(q).
\end{equation}
The first outcome of the classical polymerization (\ref{PT}) and its
associated Hamiltonian (\ref{PHamiltonian}) is that the momentum is
periodic and its changing range should be according to
$p\in[-\frac{\pi}{\lambda},+\frac{\pi}{ \lambda})$ which in the
limit $\lambda\rightarrow 0$ recovers the usual range for the
canonical momentum $p\in(-\infty,+\infty)$.

\subsection{Darboux chart}
We may consider the phase space with coordinates $(q,p)$ as a
two-dimensional symplectic manifold ${\mathcal M}$ equipped with a
closed nondegenerate $2$-form $\omega$ as its symplectic structure.
According to the Darboux theorem \cite{Arnold}, a local chart always
exists in which this $2$-form takes the canonical form

\begin{eqnarray}\label{Dar-twoform}
\omega=dq\wedge\,dp.
\end{eqnarray}
By means of the polymer corrected Hamiltonian given by the relation
(\ref{PHamiltonian}), one can write the time evolution of the system
with the help of the Hamiltonian vector field ${\bf x}_{_H}$ which
satisfies the equation

\begin{equation}\label{VF-D}
i_{\bf x}\,\omega=dH_{\lambda}.
\end{equation}
Considering the effective Hamiltonian (\ref{PHamiltonian}) by
solving above equation we reach to the equation

\begin{eqnarray}\label{Dar-VF}
{\bf x}_{_H}=\frac{\sin(\lambda p)}{m\lambda}\frac{\partial}{
\partial q}-\frac{\partial U}{\partial q}\frac{\partial}{
\partial p}.
\end{eqnarray}
Now, we are going to consider the system as a many particle system
and to apply the above formalism we need a statistical mechanics
point of view. In this regard, we recall the Liouville theorem
according to which the volume of a $2D$-dimensional symplectic
manifold is given by

\begin{eqnarray}\label{Vol-D}
\omega^{_D}=\frac{(-1)^{D(D-1)/2}}{D!}\,\underbrace{\omega\,\wedge
\,...\,\wedge\,\omega}_{D\,\,\mbox{times}}.
\end{eqnarray}
In the special case of two-dimension manifolds, this volume
coincides with the symplectic 2-form. Therefore, for a one
dimensional space with spatial volume $L$, the total volume will be

\begin{eqnarray}\label{Dar-TV}
\mbox{Vol}(\omega^1)=\int\omega^1=\int_{L}dq\times\int_{
-\frac{\pi}{\lambda}}^{+\frac{\pi}{\lambda}}dp=2\pi\Big(
\frac{L}{\lambda}\Big).
\end{eqnarray}
It is seen that in spite of the standard classical mechanics in
which the total volume of the phase space is infinite due to no
restriction on the momentum of the test particles, here the total
volume of the phase space is finite because of an upper bound for
the momenta in the classical polymeric systems. In other word, a
circle of radius $\lambda^{-1}$  determines the topology of the
momentum part of the polymeric symplectic manifold \cite{LT}.

\subsection{Noncanonical chart}
In this subsection, to study the statistical aspects of the
polymeric systems we adopt another approach by introducing the
noncanonical transformation

\begin{equation}\label{NC-Transformation}
(q,p)\rightarrow\,(q',p')=\Big(q,\frac{2}{\lambda}\sin(\frac{
\lambda p}{2})\Big),
\end{equation}
which transforms the Hamiltonian
$H_{\lambda}(q,p)\rightarrow\,H_{\lambda}(q',p')$ with

\begin{equation}\label{NC-Hamiltonian}
H_{\lambda}(q',p')=\frac{p'^2}{2m}+U(q').
\end{equation}
Although the Hamiltonian (\ref{NC-Hamiltonian}) has the standard
form like the Hamiltonians in classical mechanics, the new momentum
$p'$ is bounded as $p'\in[-\frac{2}{ \lambda},+\frac{2}{\lambda})$
due to the above transformation. In this sense the Hamiltonian
(\ref{NC-Hamiltonian}) is different from the standard classical
Hamiltonians. It is easy to see that the corresponding 2-form in
this noncanonical chart takes the form

\begin{eqnarray}\label{NC-Structure}
\omega=\frac{dq'\wedge\,dp'}{\sqrt{1-(\lambda p'/2)^2}}.
\end{eqnarray}
Now, the Poisson bracket between the variables $p'$, $q'$ is given
by

\begin{equation}\label{NC-PA}
\{q',\,p'\}=\sqrt{1-(\lambda p'/2)^2}\,,
\end{equation}
which turns to its nondeformed counterpart $\{q,\,p\}=1$ in the
limit $\lambda\rightarrow\,0$. Calculating again the Liouville
volume based on the above approach yields

\begin{eqnarray}\label{NC-TV}
\mbox{Vol}(\omega^1)=\int_{L}dq'\times\int_{-\frac{2}{
\lambda}}^{+\frac{2}{\lambda}}\frac{dp'}{\sqrt{1-( \lambda
p'/2)^2}}=2\pi\Big(\frac{L}{\lambda}\Big),\hspace{.5cm}
\end{eqnarray}which is nothing but the result obtained before in Darboux chart
(\ref{Dar-TV}). This is not an unexpected result since, the total
volume of the phase space should be invariant over the symplectic
manifold. In summary, to deal with the polymeric symplectic
manifold, one may adopt two equivalent pictures: (i) the effective
deformed Hamiltonian (\ref{PHamiltonian}), the canonical symplectic
structure in Darboux chart (\ref{Dar-twoform}) and the standard
canonical Poisson algebra; (ii) the Hamiltonian in the standard form
(\ref{NC-Hamiltonian}), the noncanonical symplectic structure
(\ref{NC-Structure}) and the nonstandard Poisson algebra
(\ref{NC-PA}). Though the trajectories on the polymeric phase space
are the same in these two pictures, as we will see in the next
section, working in the framework of the noncanonical chart is more
admissible from the statistical point of view.

\section{ S-BH and the polymeric tunneling mechanism}\label{sec3}
According to Birkhoff's theorem, the Schwarzschild metric is the
most spherically symmetric, vacuum solution of the Einstein field
equations. A Schwarzschild BH has no electric charge, $Q$, or
angular momentum, $J$. In Schwarzschild coordinates, the line
element for the Schwarzschild metric has the form

\begin{equation}\label{schw-metric}
ds^2 = -F(r) dt^2 +\frac {dr^2}{F(r)} + r^2 d\Omega^2_{2},\quad F(r)
= 1 - \frac{{2M}}{{{r}}},
\end{equation}
where $d\Omega^2_{2}$ is the line element on the 2-dimensional unit
sphere. As is well known from the standard BH thermodynamics, its
entropy $S$ is given in terms of its horizon area $A=4\pi{r^2}$ by
$S=\frac {A}{4}=\pi{r^2}$. Now, we obtain a detailed quantum
tunneling calculation from the S-BH event horizon. According to
\cite{parikh,sagh}, to describe the process of quantum tunneling
where a particle moves in dynamical geometry and crosses through the
horizon without singularity on its path, there should be a
coordinates system in terms of which the metric is not singular at
the horizon. In this sense, the above metric may be written as
\footnote{This form of the Schwarzschild metric can be obtained from
the Schwarzschild coordinates by introducing the time coordinate
\[t=t_s+2\sqrt{2Mr}+2M \ln
\frac{\sqrt{r}-\sqrt{2M}}{\sqrt{r}+\sqrt{2M}},\] where $t_s$ is
Schwarzschild time, see the first of \cite{parikh}.}
\begin{eqnarray}
ds^{2}=-\big(1-\frac{2M}{r}\big)dt^{2}+2\sqrt{\frac{2M}{r}}dtdr+dr^2+r^2d\Omega^2.
\end{eqnarray}
The metric in these new coordinates is stationary, non-static and is
regular at $r=2M$. The equation for the radial null geodesics for a
(massless) particle is given by $\dot{r}\equiv\frac{dr}{dt} =\pm
\big(1-\sqrt{\frac{2M}{r}}\big)$, in which the outgoing and ingoing
geodesics are identified by the upper and lower signs respectively.

\begin{figure}[t!]
\centering
\includegraphics{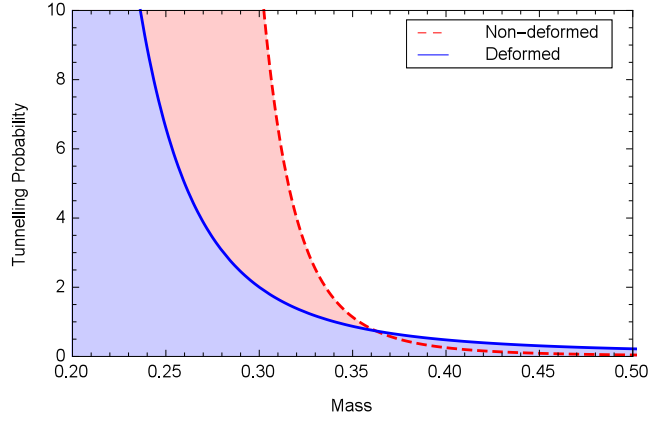}
\caption{The tunneling probabilities of S-BH in the presence of
polymer space for $\lambda=1$ and ${\cal E} = 0.1$.} \label{fig1}
\end{figure}

Here, we are going to apply the Parikh-Wilczek method to evaluate
temperature and tunneling radiation of BH in the framework of the
polymer mechanism. So, assume the trajectory of a massless particle
and consider the response of the background geometry to the radiated
quantum of energy $E$ with polymer space correction, i.e. $\cal{E}$.
Since the WKB approximation is valid near the horizon \cite{sagh},
for the classically prohibited region at the stationary phase, the
tunneling probability as a function of the particle action takes the
form
\begin{eqnarray}
\Gamma\sim\exp(-2\textmd{Im}\, {\cal{I}})\approx\exp(-E/T).
\end{eqnarray}
In order to compute $\textmd{Im}\, {\cal{I}}$, we consider the
radial null geodesics like an outgoing $s$-wave with positive energy
which passes through the horizon outwards from $r_{in}$ to
$r_{out}$. So, under condition: $r_{in}> r_{out}$ , where
$r_{in}=2M$ and $r_{out}=2(M-\cal{E})$, we have
\begin{eqnarray}
\textmd{Im}\, {\cal{I}}=\textmd{Im}\int_{r_{in}}^{r_{out}}p_rdr
=\textmd{Im}\int_{r_{in}}^{r_{out}}\int_0^{p_r}dp'_rdr.
\end{eqnarray}
After using the polymer-deformed Hamiltonian equation, we get
\begin{eqnarray}
\dot{r}=\{r,H\}=\{r,p_r\}\frac{dH}{dr}\bigg{|}_{r}.
\end{eqnarray}
Since the Hamiltonian can be written as $H=M-{\cal{E}}'$, we may use
the approximation $p^2\simeq{\cal{E}}'^2 \Rightarrow
p\simeq{\cal{E}}'$, so, we have
\begin{eqnarray}
\textmd{Im}\,{\cal{I}}&=&\textmd{Im}\int_{M}^{M-{\cal{E}}}
\int_{r_{in}}^{r_{out}}\frac{drdH}{\dot{r}}\nonumber\\
&=&\textmd{Im}\int_0^{{\cal{E}}}\int_{r_{in}}^{r_{out}}\frac{dr(-d{\cal{E}}')}{1-\sqrt{\frac{2(M-{\cal{E}}')}{r}}}.
\end{eqnarray}
Now, we incorporate quantum gravity effects encoded in the presence
of the polymer phase space which is derived from
\cite{snyder,snyder2}. To obtain the corresponding measure of
integrals, one needs the Jacobian of the transformation
(\ref{NC-Transformation}) which is given by \cite{BV}
\begin{eqnarray}
\int {{drd\cal{E}}}  \to \int {\frac{{{drd\cal{E}}}}{J(r,\cal{E})}}
= \int {\frac{{{drd\cal{E}}}}{\big( \sqrt{1-(\lambda {\cal{E}}/2)^2}
\big)}},
\end{eqnarray}
where $J(r,\cal{E})$ is the Jacobian of the transformation.
Therefore, the imaginary part of the action can be expressed as
\begin{eqnarray}
\textmd{Im}\,{\cal{I}}=\textmd{Im}\int_0^{{\cal{E}}}\int_{r_{in}}^{r_{out}}\frac{dr(-d{\cal{E}}')}
{{ \sqrt{1-(\lambda {\cal{E}}/2)^2}
}\bigg(1-\sqrt{\frac{2(M-{\cal{E}}')}{r}}\bigg )}.
\end{eqnarray}
The integral over $r$ can be evaluated by means of the residue
method. To do, one may deform the contour around the pole at the
horizon, where it lies along the line of integration and gives
$(-\pi i)$ times the residue
\begin{eqnarray}
\textmd{Im}\, {\cal{I}}=\textmd{Im}\int_0^{{\cal{E}}}\frac{2(-\pi
i)(M-{\cal{E}}')(-d{\cal{E}}')}{\sqrt{1-(\lambda {\cal{E}}/2)^2}}.
\end{eqnarray}
Now, the imaginary part of the action takes the following form
\begin{eqnarray}\label{rrat}
\textmd{Im}\, {\cal{I}}= \frac{{4\pi }}{{{\lambda ^2}}}\left( {2 -
\sqrt {4 - {\lambda ^2}{{\cal{E}}^2}}  - M\lambda \arcsin
\left[{\frac{{\lambda {\cal{E}}}}{2}} \right]} \right)+... \nonumber\\
\end{eqnarray}
After using a Taylor expansion the tunneling rate reads
\begin{eqnarray}\label{rat}
\Gamma&\sim&\exp\Bigg[\frac{{8\pi }}{{{\lambda ^2}}}(\sqrt {4 -
{\lambda ^2}{{\cal{E}}^2}}
+ M\lambda \arcsin \left[ {\frac{{\lambda {\cal{E}}}}{2}} \right] - 2)\Bigg]\nonumber\\
&=&\exp\Bigg[\left( {-8\pi M{\cal E} +4\pi {{\cal E}^2}} \right) +
\left( { - \frac{{2\pi }}{3}M{{\cal E}^3}+
\frac{{\pi {{\cal E}^4}}}{2}} \right){\lambda ^2}\Bigg]\nonumber\\
&=&\exp(\Delta S),
\end{eqnarray}
where $\Delta S$ is related to the change of Bekenstein Hawking
entropy before and after emission
\cite{sagh,ADK1,RN-tn,Akh06,Akh07}. In the string theory, it is
expected that the the emission rates from excited D-branes on the
Planck scales is related to differences between the counting of
microstates in the canonical and microcanonical ensembles. The first
term in the exponential function shows a spectrum of thermal
Boltzmannian. In addition, the existence of extra terms explains
that there is a polymer corrected non-thermal character of
radiation. The expression above includes corrections of quantum
gravity from polymer phase space which come from the deformed
algebra. Figure \ref{fig1} shows the tunneling probability of S-BH
for different values of BH mass. It is interesting to note that the
semiclassical tunneling probability for BH coincides with the
super-Planckian results \cite{sub}.

\begin{figure}[t!]
\centering
\includegraphics{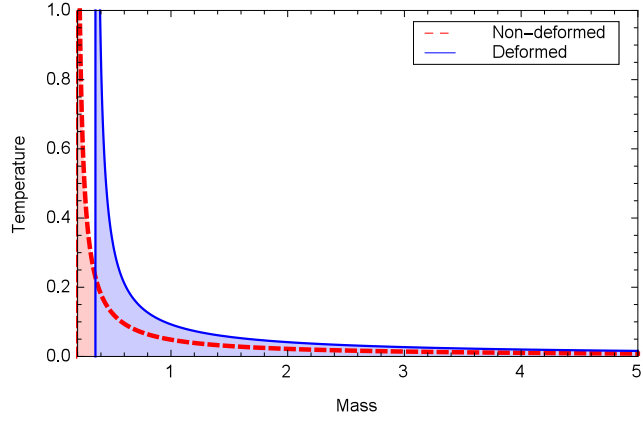}
\caption{The figure shows that the polymer-corrected of the BH's
temperature coincides with its semi-classical counterpart for large
values of the BH's mass. However, it goes to zero as the BH's mass
vanishes. The figure is plotted for $\lambda=1$ and ${\cal E} =
0.1$}. \label{fig3}
\end{figure}

In this way, the Hawing temperature for BH is given by \cite{temp}
\begin{eqnarray}
\Gamma  = \exp \left[- {\frac{{\cal E} }{T}} \right] \to T_{BH} =
\frac{{\cal E} }{2\textmd{Im}\, {\cal{I}}}.
\end{eqnarray}
Thus, we obtain the polymer corrected temperature as
\begin{eqnarray}
{T_{polymer}} = \frac{{\cal E} }{\frac{{8\pi }}{{{\lambda
^2}}}\left( {2 - \sqrt {4 - {\lambda ^2}{{\cal{E}}^2}}- M\lambda
\arcsin \left[ {\frac{{\lambda {\cal{E}}}}{2}} \right]} \right)}.
\end{eqnarray}
Now, by using of the saturated form of the uncertainty principle
${\cal E} \Delta x=\frac{1}{2}$ and $\Delta x = 2M$ \cite{temp,sagh}
(which comes from the saturated form of the Heisenberg uncertainty
principle $\Delta p\Delta x=\frac{\hbar }{2}$) in the above
equation, its Taylor expansion takes the form
\begin{eqnarray}\label{rat2}
{T_{Polymer}} = {T_{BH}}\left(\frac{{16{M^2}}}{{(8{M^2} - 1)}} +
\frac{{(16{M^2} - 3){\lambda ^2}}}{{6{{(8{M^2} - 1)}^2}}} +
...\right),
\end{eqnarray}
where ${T_{BH}}=\frac{1}{{8\pi M}}$ is the classical temperature. In
figure \ref{fig3} we have plotted the temperature in terms of mass.
As this figure shows, in the limit of large $M$, it coincides to the
Hawking temperature. However, as the mass decreases the temperature
goes to zero, a result which is in agreement, for example, with the
one-loop corrected temperature obtained in \cite{temp1}.

At this stage, let us compute the correlation between emitted
particles. When there is some correlations between different emitted
modes, some parts of the information coming out of the BH can be
retained in these correlations. We come to conclusion that the
probability of tunneling of two particles of energies $E_1$ and
$E_2$ is not equal to the probability of tunneling of one particle
with energy $E = E_1 + E_2$, i.e.

\begin{eqnarray}
\Delta S_{E_1}+\Delta S_{E_2}\neq \Delta S_{E_1+E_2}\Rightarrow \chi(E_1+E_2;E_1,E_2)\neq 0,\nonumber\\
\end{eqnarray}where $\chi(E_1+E_2;E_1,E_2)$ is the correlation
function. To see how a nonzero correlation function may come into
the play, let us compute the emission rate for a emitted quanta with
energy $E_1$ as
\begin{eqnarray}
\ln\Gamma_{E_1}= &-&8\pi M{E_1} +4 \pi {{E_1}^2}\nonumber\\
&+& \left( { - \frac{{2\pi }}{3}M{{E_1}^3} + \frac{{\pi
{{E_1}^4}}}{2}} \right){\lambda ^2}.
\end{eqnarray}
Therefore, the corresponding expression for a single quantum of
energy $E_2$, should be written as
\begin{eqnarray}
\ln\Gamma_{E_{2}}=  &-& 8\pi (M-E_1){E_2} + 4 \pi {{E_2}^2}\nonumber\\
&+& \left( { - \frac{{2\pi }}{3}(M-E_1){{E_2}^3} + \frac{{\pi
{{E_2}^4}}}{2}} \right){\lambda ^2}.
\end{eqnarray}
On the other hand, if our single quantum has the energy
$E=E_{1}+E_{2}$, its emission rate reads
\begin{eqnarray}
\ln\Gamma_{(E_1+E_2)}&=& { -8\pi M{(E_1+E_2)} +4 \pi {{(E_1+E_2)}^2}}\nonumber\\
&+& \left( { - \frac{{2\pi }}{3}M{{(E_1+E_2)}^3} + \frac{{\pi {{(E_1+E_2)}^4}}}{2}} \right){\lambda ^2}.\nonumber\\
\end{eqnarray}
Now, it is easy to show that the statistical correlation function is
non-zero and is equal to

\begin{equation}\label{kjh}
\chi\Big(E_{1}+E_{2};E_{1},E_{2}\Big)=\frac{1}{3} \pi  E_1 E_2
\left[6 E_1^2-6 M (E_1+E_2)+9 E_1 E_2+4 E_2^2\right]\lambda ^2,
\end{equation}
which includes the terms which depend on $M$ and $\lambda$. This
means that these probabilities are actually correlated. One may
compare the above mentioned correlation function with the similar
expressions obtained in \cite{ADK} and \cite{sagh,ADK1} with using
of other phenomenological approaches to quantum gravity. In each
case the correlation is appeared due to incorporation of the
characteristic parameter ($\lambda$ in our case) of the relevant
theory. To understand the role of this non-thermal correlation, note
that once the BH emits a quantum particle from its horizon, the
emission of the next quantum is affected by aberrations (on the
Planck scale) resulting from the propagation of the first quantum.
Particularly, the effects of these aberrations are more important
when the BH mass is of the order of Planck mass. In this way, the
information problem may be addressed by means of the back reaction
effects come from the quantum gravity modifications. Indeed, one may
suppose that information will leak out from the BH as a
\emph{non-thermal} correlations within the Hawking radiation.

\section{Summary}\label{sec5}
In this letter, we have studied the effects of polymer phase space
on the tunneling rate from the S-BH. After a brief review of the
polymer representation of quantum mechanics, we have introduced the
classical polymerization by means of which the Hamiltonian of the
theory under consideration gets modification in such a way that a
parameter $\lambda$, coming from polymer quantization, plays the
role of a deformation parameter. In order to apply this mechanism on
the S-BH,  we first presented the tunneling probability as a
function of the particle action. Then, we have applied the
polymerization on the model to get the polymer-modified expression
for the tunneling probability and compared the result with the
standard formalism. We have shown that in the presence of polymer
effects, tunneling mechanism of a particle is totally deviated from
thermal emission. Finally, we evaluated the correlation function
between emitted particles, and showed that the correlation is
non-zero. This result leads us to the fact that within the Hawking
radiation, there is also some information leak in the form of
non-thermal emission.

\vskip 0.2 in \textit{Acknowledgments:} The work of B. Vakili has
been supported financially by Research Institute for Astronomy and
Astrophysics of Maragha (RIAAM) under research project NO.
1/5237-111.

\end{document}